\documentclass[conference]{IEEEtran}
\IEEEoverridecommandlockouts
\usepackage{cite}
\usepackage{amsmath,amssymb,amsfonts}

\usepackage{algorithmic}
\usepackage{siunitx,booktabs}
\usepackage{xcolor}
\usepackage{colortbl}
\usepackage{graphicx}
\usepackage{amsfonts}
\usepackage{pifont}
\newcommand{\xmark}{\ding{55}}%
\usepackage{amssymb}
\usepackage{pifont}
\usepackage{multirow}
\usepackage{tabularx,ragged2e}
\usepackage{booktabs}

\definecolor{Gray}{gray}{0.83}

\newcolumntype{a}{>{\columncolor{Gray}}c}
\newcolumntype{b}{>{\columncolor{white}}c}

\usepackage{float}
\restylefloat{table}
\usepackage{color,soul}

\usepackage[hang, small, bf, margin=0pt, tableposition=top]{caption}

\usepackage{graphicx}
\usepackage[a4paper, margin=2.5cm]{geometry}
\usepackage{booktabs,subcaption,graphicx}
\def\sym#1{\ifmmode^{#1}\else\(^{#1}\)\fi}

\usepackage{textcomp}
\usepackage{float}
\usepackage{xcolor}
\def\BibTeX{{\rm B\kern-.05em{\sc i\kern-.025em b}\kern-.08em
   
T\kern-.1667em\lower.7ex\hbox{E}\kern-.125emX}}

\usepackage{hyperref}
\hypersetup{
colorlinks=true,
linkcolor=blue,
filecolor=magenta,
urlcolor=cyan,
citecolor=black,
pdfpagemode=FullScreen
}
\begin{document}
\title{Distributed Twins in Edge Computing: Blockchain and IOTA}
\author{\IEEEauthorblockN{%
Anwar Sadad \IEEEauthorrefmark{1}, Muazzam A. Khan \IEEEauthorrefmark{1}, Baraq Ghaleb 
\IEEEauthorrefmark{2}, Fadia Ali Khan 
\IEEEauthorrefmark{3}, Maha Driss\IEEEauthorrefmark{4}, 
}
\IEEEauthorblockN{Wadii Boulila\IEEEauthorrefmark{5}, Jawad Ahmad \IEEEauthorrefmark{2}}
\\
\IEEEauthorblockA{\IEEEauthorrefmark{1} Department of Computer Science, Quaid-i-Azam University, Islamabad, Pakistan.}
\IEEEauthorblockA{\IEEEauthorrefmark{2} School of Computing, Engineering \& The Built Environment,  Edinburgh Napier University, Edinburgh, UK.}
\IEEEauthorblockA{\IEEEauthorrefmark{3}Department of Mechatronics Engineering, Wah Engineering College, University of Wah, Wah, Pakistan}
\IEEEauthorblockA{\IEEEauthorrefmark{4}Security Engineering Laboratory, CCIS, Prince Sultan University, Riyadh, Saudi Arabia}
\IEEEauthorblockA{\IEEEauthorrefmark{5}Robotics and Internet-of-Things Laboratory, Prince Sultan University, Riyadh, Saudi Arabia}

}
\maketitle

\begin{abstract}
Blockchain (BC) and Information for Operational and Tactical Analysis (IOTA) are distributed ledgers that record a huge number of transactions in multiple places at the same time using decentralized databases. Both BC and IOTA facilitate Internet-of-Things (IoT) by overcoming the issues related to traditional centralized systems, such as privacy, security, resources cost, performance, and transparency. Still, IoT faces the potential challenges of real-time processing, resource management, and storage services. Edge computing (EC) has been introduced to tackle the underlying challenges of IoT by providing real-time processing, resource management, and storage services nearer to IoT devices on the network's edge. To make EC more efficient and effective, solutions using BC and IOTA have been devoted to this area. However, BC and IOTA came with their pitfalls. This survey outlines the pitfalls of BC and IOTA in EC and provides research directions to be investigated further.
\end{abstract}


\IEEEpeerreviewmaketitle   
\section{\large Introduction}
\let\thefootnote\relax\footnotetext{This work is supported by Prince
Sultan University in Saudi Arabia}

IoTs is an emerging technology capable of connecting the real world with devices that can generate or transmit data, communicate with one another, and remotely control objects via the Internet in the absence of humans. With the help of Wireless Sensors Networks and Radio Frequency Identification \cite{2}, IoT is extensively evolved in healthcare, industries, education, smart homes, smart cities, surveillance, and smart agriculture \cite{3,driss2021microservices,khan2022voting}. For data processing of IoT, cloud computing (CC) serves as a backbone due to its scalability and flexibility. However, there are significant issues associated with CC, like real-time processing, resource allocation, and security of data \cite{30}. To provide real-time data processing, EC has been introduced, which allows data processing on the edge of the network. However, issues associated with its security still exist, such as leakage of sensitive information, denial of service attacks, access control, and privacy of data.

Distributed immutable ledgers, such as BCs and IOTA, have recently shown feasibility in overcoming the aforementioned security issues of IoT. BCs use a peer-to-peer (P2P) network for communication which allows one to directly initiate payment and send it to another party without the involvement of any trusted third parties \cite{9}. In other words, it is a distributed database system linked in a P2P network through advanced cryptographic techniques that use a distributed ledger to ensure the security of messages or transactions exchanged over the network \cite{10}.

BC (e.g., bitcoin and ethereum) faces challenges when used in small day-to-day transactions (e.g., micropayments). Therefore, the research community proposed IOTA to tackle such problems \cite{11}. IOTA differs significantly from bitcoin because it is not developed based on BC technology. To facilitate low-cost micropayments, the developers of IOTA have developed an entirely different architecture using directed acyclic graph (DAG) known as tangling \cite{13}. 

To overcome the underlying challenges of EC, including security, privacy, latency, accessibility, data leakage, scalability, resource handling, throughput, reliability, control management, and energy efficiency, many contributions based on BC and IOTA have been introduced in the last few years. To the best of our knowledge there is no survey or work exist in the literature that highlights the contributions of BC and IOTA in EC. Therefore, the purpose of this study is to present a survey that overviews recently proposed methods based on BC and IOTA in EC and conclude which technology (BC or IOTA) serves best based on chosen parameters for EC.

The main contributions of the proposed work are as follows.
\begin{itemize}
  \item In literature, there is no work that discusses the contributions of BC and IOTA in EC.
  \item We have discussed that how BC and IOTA cope with underlying challenges of EC.
  \item We have comparatively analyze the role of BC and IOTA in EC.
  \item Finally, we conclude with future direction in EC using BC and IOTA.
\end{itemize}

The rest of the paper is organized as follows. Section 2 overviews both BC and IOTA and highlights contributions being made in EC using BC and IOTA. Section 3 comparatively analyzes proposed methods in terms of chosen parameters. Finally, Section 4  concludes the study with some recommendations.

\section{\large Literature Survey}
This section briefly discusses BC and IOTA, and explores the contributions being made in EC using BC and IOTA.

\subsection{\large Blockchain and IOTA}
This section introduces BC and IOTA along with their strength and weaknesses.

The BC technology, as roughly shown in Figure 1, is known for its security which is protected based on three aspects such as, decentralization \cite{21}, advanced cryptographic algorithms \cite{24}, and consensus protocols \cite{29}.

As depicted in Figure 1, each system in BC technology can be represented as nodes. Whereas, each node hold a ledger that consists of information like block number, header, timestamp, hash to previous block, own hash and the record of transactions made in the chain.

\begin{figure}
    \centering
\includegraphics[width=0.7\linewidth]{"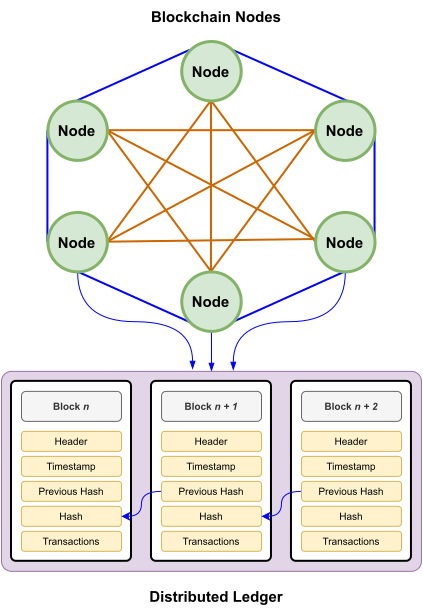"}
    \caption{Rough Presentation of BC Technology}
\end{figure}

BC uses a distributed ledger that facilitates P2P communication without a trusted third party. It is evident that BC is a secure technology based on the principles of hashing and digital signatures. However, due to the progress made in quantum computing, BC security is at risk as it has been shown recently that some of its underlying cryptographic algorithms can easily be cracked by Grover's and Shor's \cite{18} algorithms. In addition, there are various shortcomings of using BC in IoT, including the unfair/high transaction fee and the constrained nature of machine-to-machine communication. Therefore, new distributed ledger has emerged called IOTA \cite{33}, which overcomes the issues of BC concerning IoT and provide secure communication based on a concept called DAG. The underlying features provided by IOTA are scalability, zero fee transaction, quantum immune, security, and low resource requirements \cite{35}. 


\subsection{\large Blockchain in Edge Computing}
This section presents the role and contributions of BC in EC, as depicted in Table I.
The widespread success of CC is evident; however, it is not an all-in-one solution with the major issues including centralization of resources, high latency and jitter in gaming, augmented reality, and e-health \cite{19}. EC was introduced to overcome the above mentioned issues by providing computing power and storage near to edge network \cite{30}. However, EC alone cannot solve the issues regarding privacy and security of data in IoT. Hence, the focus shifted to exploring how distributed ledger-based technologies, including BC and IOTA, can be used to address such issues. Considering such importance, the authors in \cite{20} presented a survey on the integration of BC and EC showing how problems of data privacy, integrity, leakage, and access control can be solved with such integration. For secured access control, the study in \cite{21} proposed a distributed trusted authentication system using BC in EC. The study guarantees a secured access control and achieve activity traceability of terminals based on dynamic name resolution and elliptic curve cryptography. However, it stores authentication and logs data using a practical Byzantine fault tolerance consensus algorithm, compromise on authentication data may provide unintended access to the network by exploiting desired authentication or log data.\\

\def\arraystretch{1.8}%
\noindent
\begin{table*}
\label{tab:two}
\caption{Proposed solutions based on BC to tackle underlying challenges of EC}
\begin{tabular}{c|p{3.3cm}|c|c|c|c|c|c|c|c|c|c|c}
\hline
 No. & Authors  & SEC & DP & LAT & ACC & DL & SCA & RH & THR & REL & CM & EE \\
\hline
1 & Guo, S., et al \cite{21} & \checkmark & \xmark & \checkmark & \xmark & \xmark & \checkmark & \checkmark & \checkmark & \checkmark & \xmark & \checkmark \\
2 & Jayasinghe, U., et al \cite{22} & \checkmark & \checkmark & \xmark & \xmark & \xmark & \xmark & \xmark & \xmark & \checkmark & \xmark & \xmark \\
3 & Zhang, L., et al \cite{23} & \xmark & \xmark & \xmark & \xmark & \xmark & \xmark & \checkmark & \checkmark & \checkmark & \checkmark & \xmark \\
4 & Yuan, L., et al \cite{24} & \checkmark & \checkmark & \checkmark & \xmark & \checkmark & \xmark & \xmark & \checkmark & \xmark & \checkmark & \xmark \\
5 & Huang, Y., et al \cite{25} & \xmark & \xmark & \checkmark & \xmark & \xmark & \checkmark & \xmark & \checkmark & \checkmark & \checkmark & \checkmark \\
6 & Miao, Q., et al \cite{26} & \xmark & \checkmark & \checkmark & \xmark & \checkmark & \xmark & \xmark & \xmark & \checkmark & \xmark & \xmark \\
7 & Tzenetopoulos, A., et al \cite{28} & \xmark & \xmark & \checkmark & \xmark & \xmark & \xmark & \xmark & \checkmark & \checkmark & \checkmark & \checkmark \\
8 & Ahmad, A., et al \cite{29} & \checkmark & \xmark & \checkmark & \xmark & \checkmark & \xmark & \checkmark & \checkmark & \checkmark & \xmark & \checkmark \\
9 & Abdi, A.I., et al \cite{30} & \checkmark & \checkmark & \checkmark & \checkmark & \xmark & \checkmark & \xmark & \xmark & \checkmark & \checkmark & \checkmark \\
\hline

\multicolumn{13}{c}{\textbf{ABBREVATIONS}} \\ [-1ex]
\multicolumn{13}{c}{
\textbf{SEC}: Security, \textbf{DP}: Data Privacy, \textbf{LAT}: Latency, \textbf{ACC}: Accessibility, \textbf{DL}: Data Leakage, \textbf{SCA}: Scalability,
} \\ [-1ex]
\multicolumn{13}{c}{
\textbf{RH}: Resource Handling, \textbf{THR}: Throughput,  \textbf{REL}: Reliability, \textbf{CM}: Control Management,
} \\ [-1ex]
\multicolumn{13}{c}{
\textbf{EE}: Energy Efficiency
} \\ [0.2ex]
\hline
\end{tabular}
\end{table*}
To provide efficient authentication in IOTs, the authors in \cite{22} proposed a method called SCAB-IoTA based on BC which employs a secure mechanism to make clusters based on angular distance. To become a member of cluster, each IoT device is required to authenticate itself if it is in the radius of the desired cluster. However, encryption and decryption of SCAB-IoTA consume more energy and the scalability has been set aside respectively. BC restricts adversaries from alteration of data but may compromise stakeholders' data. To preserve the privacy of parties involved in chains, the study in \cite{23} presented a TrustChain solution by combining BC with trust concepts to exclude problems associated with traditional BC architectures. They defined trust as a qualitative or quantitative property of a trustee measured by a trustor for a given task in a specific context and in a specific time period. They have used a trust-based consensus management protocol to evaluate trust based on nodes' knowledge, experience, and reputation. One major issue related to such a technique is starvation because communication with the desired node depends on trust. Nodes with high trust can frequently communicate, whereas nodes with low trust will never be able to communicate with other nodes. To tackle resource allocation problem in EC, the study in \cite{24} proposed a three-tier architecture based on BC technology. It consists of a group-agent strategy with trust computing, a stacked task sorting and ranking mechanism, and a secured and efficient content model. Fake edge devices can compromise a group's trustworthiness, and tasks with low ranks may face starvation. Uploading cipher text to the cloud and indexing it using the BC may result in overhead in the case of large-scale data. Multi-access EC \cite{30} introduced an extended form of CC that allows storage services at the network edge to provide low-latency data retrieval. However, trust and incentive are two major problems in collaborative edge storage. To overcome such issues, the authors in \cite{25} have proposed novel collaborative edge storage based on BC to address incentive and trust evaluation using the historical performance of edge servers. Edge server (data off-loader) can publish a task for which other edge servers contend for the task. Reliable edge servers with a good reputation and guaranteed response time can be selected. One major issue in such an approach is when a new edge server wants to enter the ecosystem with no previous performance record. BC can also help in the efficient utilization of resources in edge environments. For the utilization problem, the authors in \cite{26} have proposed a system that can ensure fair and efficient utilization of resources on edge devices rendering it more scalable. They have proposed a data migration algorithm with consensus having low energy consumption in edge devices along with a new proof of stake mechanism. Due to high mobility in the edge environment, the nodes are moving in a small range, but the network topology remains the same. To adapt to topological changes, a migration algorithm has been designed to reallocate the data and block storage to devices dynamically. They have also used a cache mechanism to provide recent block allocation, which can reduce the overhead of missing blocks. One issue in such approach is the overflow of cache memory and other is the node having cache is down? Obviously, IoT devices are intended to share massive data with each other to impose quality-of-service \cite{driss2020servicing}. However, data sharing may result in data leakage of providers' data. Such concern is realized in \cite{27}, which presented a data sharing model based on a secured data mechanism called "BP2P-FL". It is team-based data sharing with reward and punishment mechanisms that are used to ensure data sharing with high quality and reliability. Team-based data sharing provides a collaborative environment in which a team sponsor initiates a task and assigns it to its members. Team-sponsor evaluates the capability of members based on their contribution in task completion, and members with high efforts get rewarded, whereas poor participation of members is punished by excluding them from the team. The experimental results show that the proposed method exhibited high accuracy and enhanced privacy in IoT. 

\subsection{\large IOTA in Edge Computing}
This section addresses the contributions of IOTA in EC as depicted in Table II. As illustrated in Figure 2, each system in IOTA can be presented as node. There are three phases namely, Tip, Unconfirmed, and Fully Confirmed through which each node passes. Initially, when the nodes are added it passes through Tip phase, then Confirmed and when all nodes confirm the authenticity of the added node, it goes to Fully Confirmed phase and become part of the network.
There are various challenges in integrating of BC with EC, such as scalability, latency, high energy consumption, fairness, and sensitivity to quantum computing. Therefore, IOTA, which uses DAGs, comes in to tackle issues introduced exist in BC. To address scalability in IoT, the authors in \cite{33} proposed a Scalable Distributed Intelligence Tangle-based approach to allow the integration of IoT devices across various applications. They also have presented a new proof-of-work (PoW) that enhanced energy efficiency. The experimental results show the achievement of scalability along with maintaining energy efficiency. For PoW, they have used an isolated server that performs heavy tasks to minimize energy consumption. However,  this study suffers from the isolated server's single-point-of-failure problem and the data processing management.

\begin{figure}
    \centering
   
\includegraphics[width=0.7\linewidth]{"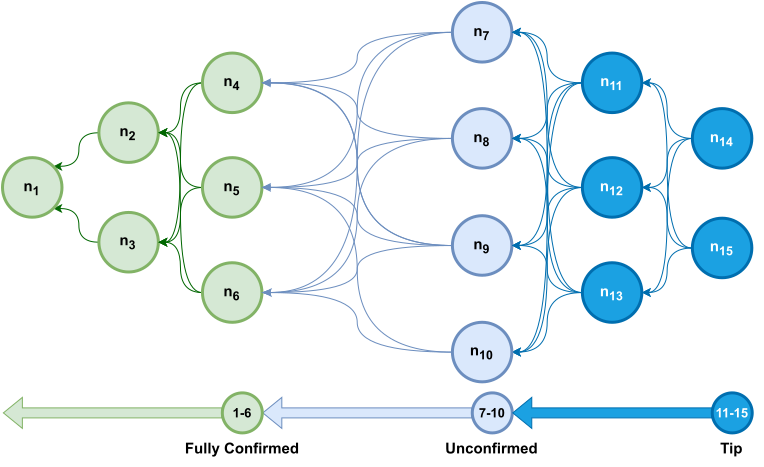"}
    \caption{Rough Presentation of IOTA Technology}
\end{figure}

For energy efficiency problem the authors in \cite{34} have proposed a novel Mobile-Agent Distributed Intelligence Tangle-Based approach to manage resources and deploy IoT applications that are both scalable and energy efficient. They have performed a PoW on IoT devices to reduce energy consumption on resource-intensive devices. The proposed approach facilitates local interaction, collection, aggregation of transactional data, and an efficient route plan. One potential problem in such an approach is the performance of PoW on dedicated servers, which may increase the cost of maintenance.  

\def\arraystretch{1.8}%
\noindent
\begin{table*}
\scriptsize
\label{tab:two}
\caption{Proposed solutions based on IOTA to tackle underlying problems}
\begin{tabular}{c|p{3.3cm}|c|c|c|c|c|c|c|c|c|c|c}
    \hline
 No. & Authors  & SEC & DP & LAT & ACC & DL & SCA & RH & THR & REL & CM & EE   \\
   \hline
1 & Alsboui, T., et al \cite{33} & \checkmark & \checkmark & \checkmark & \xmark & \xmark & \checkmark & \xmark & \xmark & \checkmark & \xmark & \checkmark \\
2 & Alsboui, T., et al \cite{34} & \checkmark & \xmark & \checkmark & \xmark & \xmark & \checkmark & \checkmark & \checkmark & \checkmark & \xmark & \checkmark \\
3 & Hellani, H., et al \cite{35} & \checkmark & \checkmark & \xmark & \xmark & \xmark & \checkmark & \checkmark & \checkmark & \checkmark & \checkmark & \xmark \\
4 & Abdullah, S., et al \cite{36} & \checkmark & \checkmark & \xmark & \checkmark & \xmark & \xmark & \xmark & \checkmark & \checkmark & \xmark & \xmark \\
5 & Fang, Z., et al \cite{37} & \checkmark & \checkmark & \xmark & \xmark & \checkmark & \xmark & \xmark & \checkmark & \checkmark & \xmark & \checkmark \\
6 & Chen, Y., et al \cite{38} & \checkmark & \xmark & \xmark & \checkmark & \checkmark & \checkmark & \xmark & \xmark & \checkmark & \xmark & \xmark \\ 7 & Carelli, A., et al \cite{39} & \checkmark & \checkmark & \xmark & \xmark & \checkmark & \checkmark & \xmark & \xmark & \checkmark & \xmark & \checkmark \\
\hline

\hline
\end{tabular}
\end{table*}
IOTA classifies nodes into full and light nodes and can manually connect light nodes with full nodes using the IOTA client balancer. This overcharges full nodes and degrades the performance of the IoT platform. To cope with such an issue, the authors in \cite{35} introduced a mechanism that fairly distributes the task among all nodes. They have developed an enhanced resource allocation algorithm called weight least connection which has improved the balancing of data traffic among full nodes based on their weights and active connections. For secure data sharing, a framework-based solution in \cite{36} has presented using masked authentication messaging (MAM) with a tangle. For the transfer of data to distributed ledgers, MAM provides an additional layer of security with cryptographic functionalities, which maintains the integrity, authenticity, and confidentiality of data. Since IoT devices are restricted in terms of memory and computing power, such a solution is quite expensive for IoT. However, the focus of security analysis in IoT is limited to the node level, while the interactive nature of the device has been ignored. To cover up such a hole, the authors in \cite{37} have proposed a framework that is used to monitor and detect potential danger to IoT devices. IOTA has been used for generating the attack graphs, probably pop-up resources, along with attack traces that can be compromised. They have also identified the dependencies between various devices by which adversaries can employ severe attacks. The proposed IOTA based model has been tested against 37 syntactic smart home systems, showing that it is highly effective and efficient. However, using such a model in edge environments will result in high energy consumption, which needs to be minimized for deployment in highly scalable networks. Furthermore, an attack called Parasite Chain Attack (PCA) \cite{40} in which the attacker invincibly builds a sub-tangle that results in double-spending. The consequences of such an attack may be catastrophic in terms of finance because it can affect the entire network by double-spending if it is not adequately prevented. Therefore, the study \cite{38} presented a scheme to tackle the PCA attack. They have proposed an algorithm for the prevention of PCA based on price splitting to slow the formation of the parasite chain effectively. However, effective detection and prevention mechanisms are required to operate in a scalable environment consuming less energy. Sensors are primary elements in IoTs, for sensors' security and privacy, the study in \cite{39} proposed L2sec, a cryptographic protocol to secure source data exchange over the IOTA. Obtained results have shown better performance in terms of effectiveness and scalability.

\section{Comparative Analysis}
This section comparatively analyzes both BC and IOTA in terms of security, privacy, latency, accessibility, data leakage, scalability, resource handling, throughput, reliability, control management, and energy efficiency. \\

The power and flexibility of IoT enable the creation of smart environments and reduce human efforts across many areas; namely, smart health, smart homes, smart cities, and smart vehicles \cite{38}. To overcome the underlying challenges of EC, the concept of distributed ledgers (BC and IOTA) has been deployed in EC \cite{29,37}. It is obvious that the integration of BC and IOTA in EC is itself a challenge that offers more issues related to scalability, energy efficiency, and security. This survey is conducted with the purpose of studying the integration effects of BC and IOTA with EC.

Solutions based on BC in EC are depicted in Figure 3, which shows the distribution of defined parameters that BC tries to tackle. Challenges offered by EC have been intelligently overcome using BC. However, BC itself has various limitations such as scalability, latency, power consumption, fee fairness, and privacy of stakeholders \cite{37}. The latest proposed papers on BC in edge networks from 2019 to 2020 have been scrutinized in which it is found that energy efficiency, scalability, and resource handling are still set aside, which are serious issues related to EC. Issues related to privacy, data leakage, accessibility, and latency have been considered, and plenty of work is devoted to scalability, energy efficiency, and scalability. BC is resource intensive and designed for large transactions, whereas IoT consists of devices with limited resources, which makes it difficult for IoT to incorporate BC efficiently. Since IoT is a scalable network that connects devices ranging from hundreds to thousands in number, the new node requires days of delay to be part of the ledger. To be integrated with EC, BC requires more intelligent and lightweight solutions that tackle scalability, energy efficiency, and security preservation.

\begin{figure}
    \centering
\includegraphics[width=0.9\linewidth]{"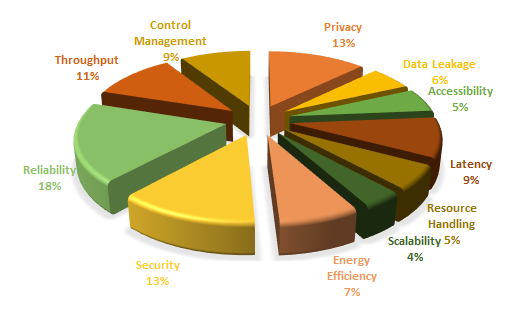"}
\caption{Distribution of proposed solutions using BC in EC}
\end{figure}

Challenges offered by BC can be resolved by using IOTA \cite{36} which is based on DAG, it is more secure and scalable than BC. It tackles scalability, fee fairness, micro-payments, and energy efficiency problems effectively \cite{28}. It is more resistant to quantum attacks as compared to BC \cite{10}. Fig 4 depicts the contributions deployed in EC in recent years, from 2020 to 2022. The use of IOTA in EC tackled scalability, accessibility, and energy efficiency problems \cite{29}. Plenty of recent works deal with privacy, data leakage, accessibility, scalability, and energy efficiency, where resource handling and latency still require a prominent solution. However, IOTA is also prone to attacks, namely conflicting transactions, blowball, lazy tips, API, and social engineering attacks (detailed study can be found in \cite{41}). Considering such attacks, more robust and resistive techniques are required to be introduced. Furthermore, to facilitate EC, IOTA based solutions are required to efficiently use energy and provide intelligent resource handling mechanisms for enhancing the utilization of resources and improving the over-all performance of the IoT.


\begin{figure}
    \centering
    \includegraphics[width=0.9\linewidth]{"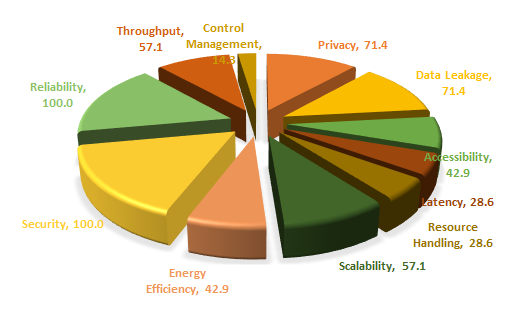"}
   
\caption{Distribution of proposed solutions using IOTA in EC}
\end{figure}

   
   

Integration of BC with EC requires prominent solutions to tackle resource handling, scalability, and energy efficiency problems. For tackling such issues, IOTA is a good option for deployment in EC. However, the latency problems need to be considered and addressed to improve the system's overall performance. Data accumulation, backup, monitoring of devices, and control management \cite{28} are also significant issues in EC that need to be addressed using BC and IOTA.

\section{ Conclusion and Future Work}
In this survey, we have studied recently proposed solutions deployed in EC based on BC and IOTA. The aim was to inspect the solutions in terms of security, privacy, latency, accessibility, data leakage, scalability, resource handling, throughput, reliability, control management, and energy efficiency. It has been found that BC comes with issues such as scalability, high transaction cost, energy inefficiency, and fee fairness when it is used in IoT. It has also been found that IOTA has more potential to address those problems using its DAG consensus mechanism. \\
Along with scalability problems, resource handling, latency, energy efficiency, data accumulation, backup, monitoring of devices, and control management are significant issues in EC that have been overlooked in recent years. Energy efficiency and resource handling in EC is a hot research area requiring intelligent solutions to use BC or IOTA. Similarly, there is no progress being made to cope with backup, monitoring, and control management problems in EC. These areas are also requiring considerable motivation toward the solutions to such problems. Another important direction for future research in EC while using distributed ledgers is to provide solutions for the detection and avoidance of potential attacks on IoT devices.


\bibliographystyle{IEEEtran}
\bibliography{ref}

\end{document}